\documentclass[pra,aps,showpacs,epsf,twocolumn]{revtex4}

\usepackage{graphicx}
\usepackage{epsfig}
\usepackage{amsmath}
\usepackage{amssymb}

\begin{document}
\title{Fermionic superfluidity with positive scattering length}
\author{Bout Marcelis and Servaas Kokkelmans}
\affiliation{Eindhoven University of Technology, P.O.~Box~513, 5600~MB  Eindhoven, The Netherlands}
\date{December 19, 2005}

\hyphenation{Fesh-bach}

\begin{abstract}
Superfluidity in an ultracold Fermi gas is usually associated with either a negative scattering length, or the presence of a two-body bound state. We show that none of these ingredients is necessary to achieve superfluidity. Using a narrow Feshbach resonance with strong repulsive background interactions, the effective interactions can be repulsive for small energies and attractive for energies around the Fermi energy, similar to the effective interactions between electrons in a metallic superconductor. This can result in BCS-type superfluidity while the scattering length is positive.
\end{abstract}


\maketitle

The crossover between Bardeen-Cooper-Schrieffer type superfluidity (BCS) and Bose-Einstein condensation (BEC) is a fundamental and challenging problem in physics. In ultracold Fermi gases, both types of superfluidity are related to the condensation of atom pairs, which can be regarded as composite bosons. A crossover between the BCS and BEC regimes, which has been predicted theoretically already several decades ago~\cite{Eagles,Leggett,NSR}, can be achieved by changing the strength of the interactions. While several experiments have studied certain aspects of this strongly interacting state~\cite{regal,zwierlein2,bartenstein,kinast}, only recently the existence of superfluidity in the crossover regime has been unambiguously demonstrated by experiment~\cite{zwierlein}.

Experimentally the crossover is realized by changing the strength of the two-body interactions via an s-wave Feshbach resonance~\cite{feshbach,Moerdijk}. A measure for this interaction strength at zero collision energy is provided by the scattering length $a$, which is the first parameter in the effective range expansion of the scattering phase shift~\cite{taylor}. It has been suggested that the physical properties in the crossover only depend on the parameter $(k_F a)^{-1}$, with $k_F$ the Fermi wavenumber. This has the strong implication that exactly on resonance, where $a$ goes through infinity, all thermodynamic and superfluid properties are independent of the microscopic details of the interactions, and therefore should be universal quantities for all atomic systems under study~\cite{Heiselberg,Carlson,ho,ohara}.

However, the microscopic physics that gives rise to the Feshbach resonance reveals that more than one parameter is needed to describe the scattering phase shift. Apart from the scattering length, the width of the Feshbach resonance is the most important parameter. In order to understand the origin of this width, we have to separate the resonance from the background interactions. At large relative distance, the two interacting atoms are sensitive only to the background interaction potential. At short distance, this potential is coupled to another potential with a different internal spin configuration. The latter potential is energetically closed at long distance, and exhibits a molecular state close to the collision threshold~\cite{feshbach,Moerdijk}. The coupling strength between these two potentials can be related to an energy width which, in comparison to the Fermi energy of the system, categorizes a Feshbach resonance in terms of its broadness. This width can be converted into an effective range coefficient, which is the second parameter in the effective range expansion~\cite{taylor}. Using this approach, it has been pointed out in several papers that universality does not hold for narrow Feshbach resonances~\cite{combescot,palo,bruun,petrov}.

It should be noted that in these papers the background interactions were taken to be small, and the Feshbach resonance was described using an effective range coefficient independent of magnetic field or momentum. However, in case the background interactions do have a resonant character -- which is indicated by a large background scattering length, as for example in some resonances in $^6$Li, $^{133}$Cs, and $^{85}$Rb -- the scattering phase shift shows non-trivial energy dependence beyond the effective range approximation, due to the interplay between the background and Feshbach resonance~\cite{marcelis}.

In this paper we investigate the BCS-BEC crossover for a Feshbach resonance which allows for a large background scattering length. We show that we can mimic a situation well known from condensed-matter physics, where Cooper pairs can be formed due to an effectively attractive interaction for relative wavenumbers around $k_F$, while for smaller wavenumbers the interaction is effectively repulsive. For electrons in a metal such an interaction exists due to the combined effect of the Coulomb and electron-phonon interactions~\cite{fetter}, while in atomic systems this can be achieved by the interplay of a Feshbach and a background resonance.

\section{Feshbach resonances with large background scattering length}
Feshbach resonances are presently an indispensable tool to control the atomic interactions in cold atomic physics. By simply changing the magnetic field the interactions can be tuned from weak to strong, and from effectively attractive to repulsive. We consider a two-component Fermi gas in the dilute limit, i.e. the density $n$ satisfies $nr_0^3\ll 1$, where $r_0 \simeq (m C_6/16 \hbar^2)^{1/4}$~\cite{gribakin} is the range of the potential, with $m$ the atomic mass and $C_6$ the van der Waals coefficient. The Fermi wavenumber $k_F$ is related to the density of a single spin component as $n_{\uparrow}=n_{\downarrow}\equiv n=k_F^3/6\pi^2$.

Feshbach resonances originate from the coupling of scattering atoms in an energetically open channel, to one (or several) bound state(s) in energetically closed channels. When the energy of a closed-channel bound state is close to the energy of the scattering  atoms, the scattering process becomes resonant. The effect of two-body interactions on the atomic wave functions can be summarized by the energy-dependent scattering phase shift $\delta(k)$, where $k=\sqrt{mE/\hbar^2}$, with $E$ the relative collision energy. The scattering matrix is related to the phase shift by $S(k)=\exp{2i\delta(k)}$, and for the cold alkali gases under consideration its most general form is given by \cite{marcelis}
\begin{equation}
    S(k) =  S_P(k) \frac{E - \nu_{\rm bare} - \Delta(E)-\frac{i}{2}\Gamma(E)}{E - \nu_{\rm bare} - \Delta(E)+\frac{i}{2}\Gamma(E)}. \label{smatrix}
\end{equation}
The direct part of the scattering matrix $S_P(k)$ describes the interactions in the open channel, and $\Delta(E)-\frac{i}{2}\Gamma(E)$ is the complex energy shift, which describes the dressing of the closed-channel bound state with bare detuning $\nu_{\rm bare}$ due to the coupling with the open-channel atoms.

In scattering systems with a background scattering length $a_{\rm bg}\simeq r_0$, the direct interactions are non-resonant, and the energy shift takes the simple form $\Delta-iCk$. The bare detuning is dressed by the constant term $\Delta$ such that the dressed detuning is given by $\nu=\nu_{\rm bare} +\Delta$. The scattering length diverges for $\nu=0$. The term $iCk$ describes the finite lifetime of the bound state coupled to the continuum, where the coupling strength $C$ is a constant related to the magnetic field width of the resonance (see below).

In scattering systems with a background scattering length $|a_{\rm bg}|\gg r_0$, the open channel in general has a bound or virtual (nearly-bound) state close to the zero-energy collision threshold. In this case, the resonant energy dependence of the open-channel propagator gives rise to non-trivial energy dependence of the energy shift. Explicitly taking into account the resonance poles of both the closed-channel bound state and the open-channel bound (or virtual) state, the energy shift can be written as $\Delta(E) - \frac{i}{2}\Gamma(E) = \Delta-iCk\zeta^{-1}_k$~\cite{marcelis}. Here the complex energy shift is written as its value at zero collision energy $\Delta \equiv \Delta(E=0)$, which is real, and an additional term that accounts for the energy dependence of both the real part of the complex energy shift $\Delta(E)-\Delta$, as well as the imaginary part $\Gamma(E)$. The coupling strength $C$ is related to the magnetic field width $\Delta B=B_0'-B_0$ as $C=\Delta\mu^{\rm mag}\Delta Ba^P$, $\nu=\Delta \mu^{\rm mag} (B-B_0)$ is the dressed detuning, $\Delta \mu^{\rm mag}$ is the magnetic moment difference between the open and closed channel, and $\zeta_k=1+ika^P$ with $a^P=a_{\rm bg}-r_0$ describes the additional energy dependence related to the resonant direct interactions. $B_0$ is the field where $a(B)$ diverges, and $B_0'$ is the field where $a(B)=r_0$. The interplay between the two resonances is thus described by the scattering phase shift:
\begin{equation}\label{phaseshift}
\delta(k) = -k r_0-\arg{\zeta_k}-\arg{(E-\nu + iCk\zeta^{-1}_k)}.
\end{equation}
From Eq.~(\ref{phaseshift}) it follows that the total scattering length is related to these parameters by
\begin{equation} \label{atotal}
a(B)=r_0+a^P\left(1-\frac{\Delta B}{B-B_0}\right).
\end{equation}
In the following, we will discuss scattering systems with a resonant direct interaction between the fermions. In this case the non-resonant interaction terms related to $r_0$ will generally have a small effect compared to the resonant terms, and will be ignored in the remainder of this paper.

The system can be described equivalently by taking into account the scattering poles of the $T$ matrix of both the molecular and background bound (or virtual) states~\cite{taylor}, which can be shown to correspond to solutions of
\begin{equation}
1/a + ik - R_e(k,B) k^2=0, \label{effrangeeq}
\end{equation}
where $R_e(k,B)=\hbar^2(1+ika^P)/(ma^P \Delta \mu^{\rm mag}(B-B_0'))$ is a generalized energy and field dependent effective range parameter. These poles determine the two-body $T$ matrix and scattering phase shift uniquely. In contrast to previous work~\cite{combescot,palo,bruun,petrov}, where $R_e$ was taken to be a constant, here $R_e$ depends on \emph{both} energy and magnetic field and describes the most general Feshbach resonance that can be found in ultracold alkali gases. Note that for magnetic fields where $a \simeq r_0$, the effective range parameter $R_e$ actually diverges and cannot be neglected even in the case of a broad resonance. The usually considered expressions for Feshbach resonance scattering without an open-channel resonance~\cite{Moerdijk} can be seen as a limit of the general expression Eq.~(\ref{phaseshift}) for small $a_{\rm bg}$.

\section{Many-body description}
In order to describe the many-body properties of the system, we use the thermodynamic approach introduced by Nozi\`eres and Schmitt-Rink (NSR)~\cite{NSR}. This approach is based on a simplified description of the full crossover problem and cannot be expected to give quantitatively precise results. Therefore, in recent years several directions have been pursued in order to formulate a more complete crossover theory~\cite{Strinati,holland,Ohashi,levin,stoof,giorgini,liu}. However, the NSR description does capture the essential physics giving rise to the crossover phenomenon. Especially the very simple relation between the two-body phase shifts in vacuum, and the in-medium phase shift used in NSR make this approach a very convenient framework to study the role of energy dependent scattering in a many-body system. In the following, we show how to generalize this approach in order to include the energy-dependent phase-shift given by Eq.~(\ref{phaseshift}).

Within the NSR approach, the two-body interactions are taken into account in the ladder approximation. The two-body interactions are described by the many-body $T$ matrix, which can be found by solving the Lippmann-Schwinger equation (LS)
\begin{equation}
\tilde T = U - U\chi \tilde T,
\end{equation}
where $U$ is the two-body interaction potential, that generally depends on the relative collision energy $E$, and $\chi$ is the two-particle propagator. In a vacuum, the many-body $T$ matrix should reduce to the two-body $T$ matrix, which is known analytically and gives rise to the scattering phase shift given by Eq.~(\ref{phaseshift}). The bound states and resonances are described by the poles of the two-body $T$ matrix, and uniquely characterize the scattering properties in the two-body sector. We are thus faced with constructing a potential $U$ that gives the correct poles in the two-body limit. Using this potential ensures that the two-body physics is built correctly into the many-body theory.

It can be shown from coupled-channels scattering theory that the effective interaction potential for the fermions in the open channel is given by
\begin{equation}
U_{\bf kk'}(E)=V_{\bf kk'}+\frac{g_{\bf k}g_{\bf k'}}{E-\nu_{\rm bare}},
\end{equation}
where $V$ is related to the direct interactions between fermions in the open channel, and the second term describes the coupling of fermions to and from the closed-channel bound state~\cite{noteonU}. To simplify the equations, it is convenient to replace the microscopic finite-range interaction potentials with contact potentials that correctly reproduce the scattering properties described by Eq.~(\ref{phaseshift}) in the two-body limit. However, since contact potentials do not explicitly depend on Fourier momentum, they introduce unphysical ultraviolet divergences which have to be regulated explicitly by renormalizing the potential and two-body propagator. The appropriate regularization procedure has been discussed in detail in Ref.~\cite{kokkelmans02}, where it was shown that the $T$ matrix is described by the renormalized LS equation
\begin{equation} \label{Tmatrixeff}
T(E) = U_{\rm eff}(E) - U_{\rm eff}(E) \bar\chi_0(E^+) T(E),
\end{equation}
where the renormalized effective interaction is related to the microscopic parameters of section I by
\begin{equation} \label{effint}
U_{\rm eff}(E)=\frac{4\pi\hbar^2}{m} \left(a^P + \frac{C}{E-\nu} \right),
\end{equation}
and the two-body pair-propagator in vacuum is explicitly regularized to account for the ultraviolet divergences:
\begin{equation}
\bar\chi_0(E^+) = \sum_{\bf k} \left( \frac{1}{2\epsilon_{k}-E^+} -\frac{1}{2\epsilon_{k}}\right),
\end{equation}
with $E^+$ approaching the real axis from above, and $\epsilon_k=\hbar^2k^2/2m$. It can be shown that Eq.~(\ref{Tmatrixeff}) with the effective potential $U_{\rm eff}(E)$ Eq.~(\ref{effint}) exactly reproduces the pole equation of Eq.~(\ref{effrangeeq}).

The many-body $T$ matrix is obtained by replacing the regulated pair-propagator in vacuum by the regulated in-medium two-particle propagator, which is defined as
\begin{equation}
\chi\left( {\bf q},\epsilon^+ \right)=\sum_{\bf k}\left(\frac{1- f(\xi_+)- f(\xi_-) }{ 2\epsilon_k-2\epsilon_p }-\frac{1}{2\epsilon_k} \right),
\end{equation}
where the relative energy of the atoms $E=2\epsilon_p$ defines a relative momentum $p\equiv\frac{\sqrt{m}}{\hbar}\left( \epsilon^+-\frac{1}{2}\epsilon_q+2\mu \right)^{1/2}$, with the total energy of the scattering atoms (relative to the chemical potential) $\epsilon^+$ approaching the real axis from above, and $\frac{1}{2}\epsilon_q$ the center-of-mass momentum of the scattering particles. The medium effects are described by the Fermi distribution functions
\begin{equation}
f(\xi_\pm) = \frac{1}{e^{\beta(\xi_\pm-\mu)}+1 },
\end{equation}
with $\beta$ the inverse temperature and $\xi_\pm=\epsilon_{\rm q/2 \pm k}$. The interactions are thus described by the many-body $T$ matrix
\begin{equation} \label{Tmatrix}
\tilde T\left( {\bf q},\epsilon^+ \right)^{-1} = U_{\rm eff}(2\epsilon_p)^{-1}+\chi\left( {\bf q},\epsilon^+ \right).
\end{equation}
This way of replacing the bare interactions by physical (renormalized) interactions is quite general, and closely related to a method due to Gorkov \cite{Gorkov61}, and used subsequently in different forms \cite{Haussmann93,Jensen04}.

We now proceed to give a description of the many-body system in the NSR approach. The number density $N/V=2n$, with $N$ the total number of atoms in a volume $V$, is found at a temperature $T$ (in the normal state) by first calculating the thermodynamic grand potential $\Omega$ within the ladder approximation, and then taking the derivative to the chemical potential $\mu$: $2n(\mu,T) = -\frac{\partial}{\partial \mu}\frac{\Omega}{V}$. The grand potential is then found to be
\begin{equation} \label{Omega}
\Omega = \Omega_0 + \frac{1}{\pi}\sum_{{\bf q},\epsilon} g(\epsilon) \phi({\bf q},\epsilon),
\end{equation}
where $\Omega_0$ corresponds to a non-interacting gas of fermions, $g(\epsilon)=\left( \exp(\beta\epsilon) -1 \right)^{-1}$ is the Bose distribution function, and the phase of the many-body $T$ matrix $\phi=\phi^P+\phi^Q$ has a contribution from the resonant background
\begin{equation}
\phi^P({\bf q},\epsilon) ={\rm Im}\log\left( 1+\frac{4\pi\hbar^2a^P}{m} \chi\left( {\bf q},\epsilon^+ \right) \right),
\end{equation}
as well as from the Feshbach resonance
\begin{equation}
\phi^Q({\bf q},\epsilon) ={\rm Im}\log\left( \nu-2\epsilon_p-\frac{4\pi\hbar^2C}{m} \Pi\left( {\bf q},\epsilon^+ \right) \right),
\end{equation}
where $\Pi\left( {\bf q},\epsilon^+ \right)=\chi\left( {\bf q},\epsilon^+ \right) \left( 1+\frac{4\pi\hbar^2a^P}{m} \chi\left( {\bf q},\epsilon^+ \right) \right)^{-1}$ includes the dressing of the Feshbach molecules due to the resonant background interactions. Note that we exclude the contribution from a possible pole associated with the background interactions when $a^P>0$ in the energy integral in Eq.~(\ref{Omega}), since we assume that the corresponding molecular state cannot be populated. The critical temperature is determined from the gap equation $\tilde T^{-1}({\bf q=0},\epsilon=0)=0$, which relates the instability of the gas towards the formation of condensed pairs to a divergence in the many-body $T$ matrix. Solving the number and gap equations self-consistently gives the critical temperature $T_c$ and chemical potential $\mu_c$ at this temperature as a function of the interaction parameters.

\begin{figure}[t]
\includegraphics[width=\columnwidth]{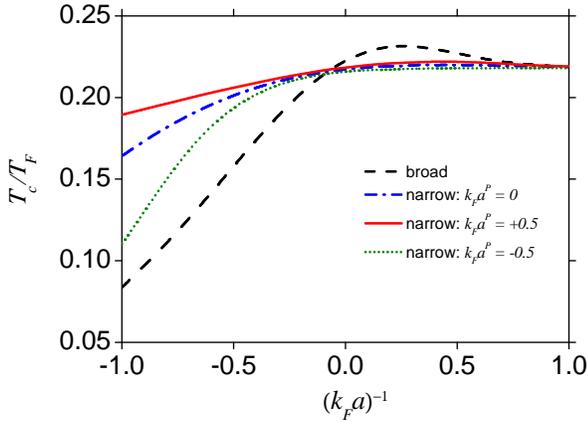} \caption{(Color Online) Critical temperature $T_c$ in the crossover region as a function of $(k_Fa)^{-1}$ in the case of broad and narrow Feshbach resonances with different background interaction strengths $k_Fa^P$. Dashed line: broad resonance where $(k_Fa)^{-1}$ is the only relevant parameter; solid, dash-dotted and dotted lines: narrow resonance where $\tilde{C}=1$ and $k_Fa^P=+0.5,0,-0.5$ respectively. \label{Fig1}}
\end{figure}

In the vacuum limit, where the two-particle propagator reduces to $imp/4\pi\hbar^2$ and the energy of the scattering particles is $\epsilon=2\epsilon_k+\frac{1}{2}\epsilon_q-2\mu$, the phase shift $\phi({\bf q},\epsilon)$ only depends on the relative energy $2\epsilon_k$ and is related to the scattering phase shift Eq.~(\ref{phaseshift}) by $\phi({\bf q},2\epsilon_k+\frac{1}{2}\epsilon_q-2\mu)=-\delta(k)-\pi$. Since $k_Fr_0 \ll 1$, we neglect the range of the background potential, and then the effective interaction Eq.~(\ref{effint}) exactly reproduces the non-trivial energy dependence of Eq.~(\ref{phaseshift}).

\section{Results and discussion}
We now turn to the numerical results. We first consider the case of a broad Feshbach resonance, characterized by $\tilde{C} \equiv 4k_FC/3\pi\mu_F \gg 1$ or equivalently $|k_F R_e(0,B_0)|\ll 1$. We solved the number and gap equations numerically with $\tilde{C} = 100$, and for different values of $-1<k_F a^P<1$. The results are not sensitive to the actual value of $k_F a^P$, and an example is shown in Fig.~\ref{Fig1} (dashed line). This reproduces the well known results previously obtained by NSR and others~\cite{NSR,Randeria}. In this case the only relevant parameter in the crossover region is the total scattering length Eq.~(\ref{atotal}), and details of the interactions at higher energy scales do not enter the phase integral Eq.~(\ref{Omega}) as they are cut off by the distribution functions.

\begin{figure}[t]
\includegraphics[width=\columnwidth]{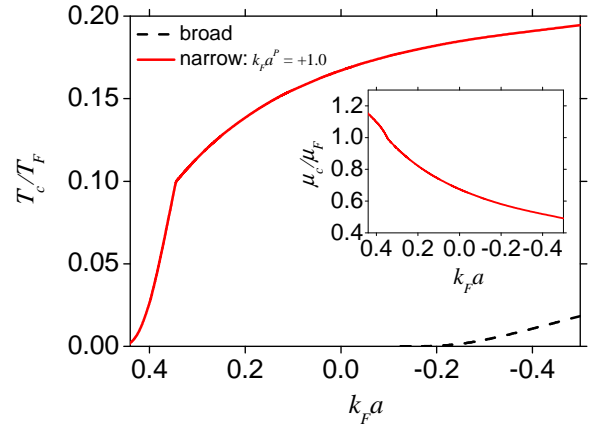} \caption{(Color online) Critical temperature $T_c$ in the BCS limit, for small negative and positive scattering length $k_Fa$, in the case of a narrow Feshbach resonance with $k_Fa^P=+1$ (solid line), and of a broad resonance (dashed line). Inset: chemical potential $\mu_c$ at $T_c$ for identical parameters. \label{Fig2}}
\end{figure}

In the case of a narrow Feshbach resonance ($\tilde{C} \lesssim 1$), details of the interactions turn out to be essential in the crossover region, as well as in the BCS- and BEC-limits. In Fig.~\ref{Fig1} the critical temperature in the crossover region is shown for several narrow resonances ($\tilde{C} = 1$) with different values of the background interaction strength $k_F a^P$ (solid, dash-dotted and dotted lines). In the BEC limit (right side of figure), the critical temperature approaches the limiting value $T_c \simeq  0.218 T_F$, and superfluidity is dominated by the condensation of molecules. Compared to a broad resonance, $T_c$ is suppressed in the crossover region, while in the BCS-limit (left side of figure), the critical temperature remains a much larger fraction of the Fermi temperature $T_F$ compared to the broad resonance case. Somewhat surprisingly, $T_c$ becomes larger in the BCS-limit if the background interactions are more repulsive.

\begin{figure}[t]
\includegraphics[width=\columnwidth]{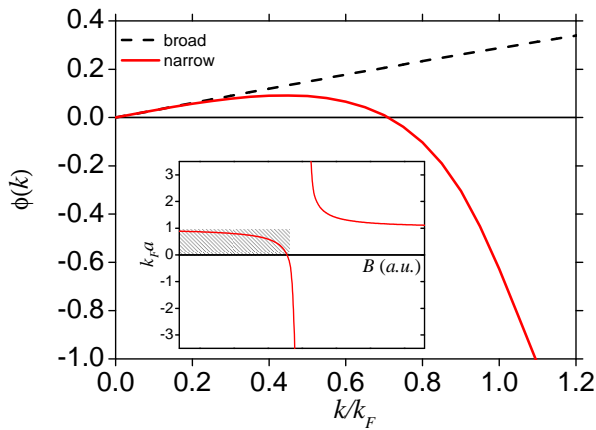} \caption{(Color online) Phase shift $\phi(k)\equiv\phi({\bf q},2\epsilon_k+\frac{1}{2}\epsilon_q-2\mu)$ for $k_Fa=+0.3$. The resonance parameters are identical to those of Fig.~\ref{Fig2}. Inset: scattering length as a function of magnetic field. The shaded area indicates the region where $a>0$, but there are no molecules in the system. \label{Fig3}}
\end{figure}

This leads us to the very interesting situation when the background scattering length is large and positive, while the Feshbach resonance is narrow. In Fig.~\ref{Fig2} the critical temperature for a narrow resonance ($\tilde{C} = 1$) with strong repulsive background interactions ($k_Fa^P=+1$) is shown (solid line) \cite{notefig2}. The cusp around $T_c/T_F=0.1$ is not physical, but an artifact resulting from the NSR approach \cite{notefig2,pieri}. If the system is initially on the BCS-side of the resonance ($k_F a\simeq -0.5$), and subsequently tuned away from resonance towards the point where $a$ becomes close to zero, we find that the critical temperature can still be a large fraction of the Fermi temperature. This in contrast with a broad resonance (dashed line), where the superfluid state is lost before $a$ changes sign. Going through the point $a=0$, the low-energy interactions become effectively repulsive as $a>0$. However, the frequency dependence of the phase shift can still give rise to attractive interactions close to the Fermi energy. In Fig.~\ref{Fig3} the phase $\phi(k)$ is shown for the same resonance parameters as in Fig.~\ref{Fig2}, and with the total scattering length equal to $k_Fa=+0.3$. A positive phase corresponds to effectively repulsive interactions, while a negative phase corresponds to attractive interactions. In the case of a broad resonance, we immediately see that for $a>0$ the interactions are repulsive at all relevant energies. For the narrow resonance with strong repulsive background interactions, although the interactions are repulsive for low collision energies, the atoms colliding at energies close to the Fermi energy feel an attractive interaction, which is the reason that the system can still become superfluid.

From the inset of Fig.~\ref{Fig2} we see that for $k_Fa\gtrsim 0.35$ the chemical potential becomes larger than $\mu_F$, reflecting the fact that superfluidity can be retained even when the total system acts as effectively repulsive. The repulsive behavior is indeed expected from the positive scattering length for the fermions. The fact that this system with repulsive interactions can still form a superfluid is a direct consequence of the energy dependence of the interactions for a narrow Feshbach resonance.

We have done an independent check of our calculation by comparing this interesting regime to the regular BCS limit. Since we are in the regime where $T_c/T_F<0.1$ and $\mu\simeq\mu_F$, the BCS result can be expected to become valid again.  We compared the functional dependence of $T_c/T_F$ versus $k_F a$ by using the well-known BCS formula
\begin{equation}
T_c/T_F=\frac{\gamma}{\pi} \exp\left(-\frac{\pi}{2 |k_F a(p)| }\right). \label{bcs}
\end{equation}
Instead of using the scattering length $a$, we replace it by a 'energy-dependent scattering length' $a(p)$, related to the relative collision energy via $p=\sqrt{m E/\hbar^2}$, defined from the scattering phase shift Eq.~(1):
\begin{equation}
a(p)\equiv-\tan\left[\delta(p)\right]/p.
\end{equation}
We probe this redefined scattering length at relative wavenumber $p=\sqrt{2 m \mu/\hbar^2}$, which corresponds to the relative momentum of two particles on opposite sides of the Fermi sphere. In the regime under consideration (using the chemical potential given in Fig.2) the value of $a(p)$ will be negative, and we find from Eq.~(\ref{bcs}) basically the same ratio of $T_c/T_F$ as in Fig.~2. By probing the energy-dependent interaction at the Fermi sphere, we thus find in the BCS limit a result that compares excellently with our enhanced NSR method.

We have studied the presently known Feshbach resonances for $^6$Li and $^{40}$K, which are the atoms currently used to study the BCS-BEC crossover. However, none of these resonances satisfy the conditions needed to observe the results presented above. But since there are several bosonic systems which do satisfy the requirement of having a narrow resonance with large $a_{\rm bg}$, there is no principal reason why a fermionic system could not be in a similar situation. Possible candidates will be fermionic alkali mixtures, such as $^6$Li$-^{40}$K.

In conclusion, a positive scattering length is usually associated with repulsive interactions in the context of ultracold atomic gases. In the case of a narrow Feshbach resonance, the interactions are strongly energy dependent. We have shown that in the special case of a large and positive background scattering length, the interactions can be repulsive for small energies, whereas attractive for energies around the Fermi energy, similar to the effective interactions between electrons in a metallic superconductor. This results in fermionic superfluidity while the scattering length is positive, without the presence of bosonic molecules in the system.

This work was supported by the Netherlands Organization for Scientific Research (NWO).


\begin{thebibliography}{99}
\bibitem{noteonU} Alternatively, this effective potential for the fermions is obtained within the framework of resonance superfluidity theory \cite{holland} by integrating out the bosons explicitly.
\bibitem{notefig2} In NSR, the use of non-interacting distribution functions $n_0(\mu)$, leads to an unphysically large $\Omega$ if $\mu>\mu_0$, where $\mu_0=\mu_0(T)$ is the chemical potential for an ideal gas. We therefore use $n_0(\mu_0)$ to find $\Omega$, in the case $\mu>\mu_0$.
\end{thebibliography}
\end{document}